\documentclass[12pt]{iopart}
\usepackage{graphicx}

\begin{document}
\letter{Polar type density of states in non-unitary odd-parity 
superconducting states of gap with point nodes
}

\author{K Miyake}

\address{Division of Materials Physics, Department of Physical Science, 
Graduate School of Engineering Science, 
Osaka University, Toyonaka, Osaka 560-8531, Japan
}

\ead{miyake@mp.es.osaka-u.ac.jp}

\begin{abstract}
It is shown that the density of states (DOS) proportional to the excitation 
energy, the so-called polar like DOS, can arise in the odd-parity states 
with the superconducting gap vanishing at points even if the spin-orbit 
interaction for Cooper pairing is strong enough.  
Such gap stuructures are realized in the non-unitary 
states, F$_{1{\rm u}}(1,{\rm i},0)$, 
F$_{1{\rm u}}(1,\varepsilon,\varepsilon^{2})$, 
and F$_{2{\rm u}}(1,{\rm i},0)$, classified by Volovik and Gorkov, 
{\it Sov. Phys.$-$JETP} {\bf 61} (1985) 843.  This is due to the fact that 
the gap vanishes in quadratic manner around the point on the Fermi surface.  
It is also shown that the region of quadratic energy dependence of DOS, 
in the state F$_{2{\rm u}}(1,\varepsilon,\varepsilon^{2})$, is restricted 
in very small energy region making it difficult to distinguish from the 
polar-like DOS.  
\end{abstract}

\submitto{JPCM}
\pacs{71.27.+a, 74.20.Rp, 74.70.Tx}

\maketitle

In early stage of research of the heavy fermion superconductors, 
it has been important to infer the gap anisotropy from the power-law of 
the temperature dependence of a series of physical 
quantities~\cite{SRMV,Miyake,HVW,EHW}.  
It was a sort of the golden rule there that point node(s) of the 
superconducting gap leads 
to the density of states (DOS), $N_{\rm s}(\omega)\propto \omega^{2}$, 
while line node(s) leads to $N_{\rm s}(\omega)\propto \omega$.  
It was also emphasized that all the odd-parity pairings would have only 
point node(s) if the spin-orbit coupling for the pairing interaction were 
so strong that the spin- and orbital degrees of freedom of the gap function 
cannot change independently~\cite{Anderson,VG,UR,Blount,SU}.  However, 
it is not so self-evident whether the spin-orbit coupling for pairing is 
really so strong to quench technically the independent variation in spin- 
and orbital space~\cite{Miyake2,Ozaki}.  In any case, the classification 
scheme proposed by Volovik and Gorkov (VG) has been believed to rule out the 
polar-like DOS for the odd-parity states.  The purpose of this letter is to 
point out that three of the non-unitary states in VG scheme has the polar-like 
DOS because the $k$-dependence of the gap around the point node is quadratic 
rather than linear.  

In the odd-parity manifold, the quasiparticle energy is the matrix in the 
representation of spin eigen state as follows~\cite{Leggett}:
\begin{equation}
{\hat E}_{\bf k}=[\xi_{\bf k}^{2}{\hat 1}+
{\hat{\Delta}}^{\dagger}_{\bf k}{\hat{\Delta}}_{\bf k}]^{1/2},
\label{dispersion}
\end{equation}
where the superconducting gap is also the 2$\times$2 matrix in the 
presentation of spin eigenstate, and is represented in terms of 
the $d$-vector as 
\begin{equation}
{\hat {\Delta}}_{\bf k}={\rm i}\sum_{j}\Delta(\sigma_{j}\sigma_{y})
d_{j}({\bf k}),
\label{dispersion}
\end{equation}
where $\sigma_{j}$ is the Pauli matrix of the $j$-the component, 
with $j$=$x$, $y$, and $z$.  
The eigen values of the magnitude of the gap matrix is given as~\cite{Leggett} 
\begin{equation}
({\hat{\Delta}}^{\dagger}_{\bf k}{\hat{\Delta}}_{\bf k})_{\pm}=
\Delta^{2}\left[({\bf d}({\bf k})\cdot{\bf d}^{*}({\bf k}))
\pm|{\bf d}({\bf k})\times{\bf d}^{*}({\bf k})|\right]. 
\label{eigenvalue}
\end{equation}
It is remarked that the time reversal symmetry is broken in the 
non-unitary state where 
${\rm i}{\bf d}({\bf k})\times{\bf d}^{*}({\bf k})\not=0$
The DOS in the superconducting state $N_{\rm s}(\omega)$ is 
expressed as follows: 
\begin{equation}
{N_{\rm s}(\omega)\over N(0)}={\omega\over 2}\sum_{\alpha=\pm}
\int{{\rm d}{\hat{\bf k}}\over 4\pi}
{\theta(\omega-|\Delta_{\alpha}({\hat {\bf k}})|)\over\sqrt{\omega^{2}-
({\hat{\Delta}}^{\dagger}_{\bf k}{\hat{\Delta}}_{\bf k})_{\alpha}}}, 
\label{dos}
\end{equation}
where $\theta(x)$ is the Heaviside function.  

\begin{figure}
\begin{center}
\includegraphics[width=0.6\linewidth]{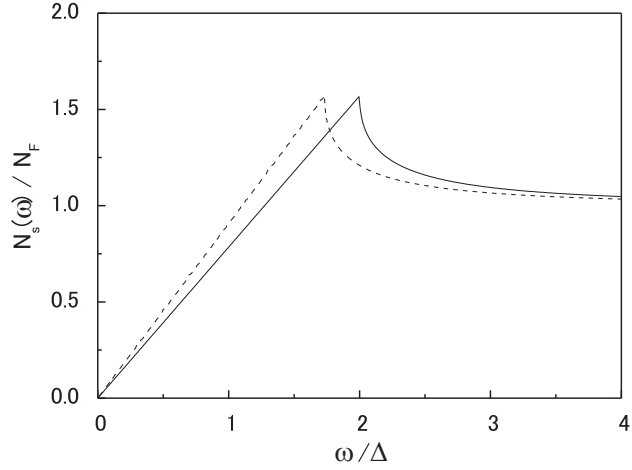}
\caption{
$N_{\rm s}(\omega)/N_{\rm F}$ in the states F$_{1u}(1,{\rm i},0)$ and 
F$_{2u}(1,{\rm i},0)$, $N_{\rm F}$ being the DOS in the normal state at the 
Fermi level.  The dashed curve is for the DOS of 
the polar state $\Delta_{\bf k}=\sqrt{3}\Delta{\hat k}_{z}$.  
}
\label{fig:1}
\end{center}
\end{figure}

The $d$-vector of the state F$_{1u}(1,{\rm i},0)$, the class of group 
theoretical representaion $D_{4}(E)$, is given as~\cite{VG} 
\begin{equation}
{\bf d}({\bf k})=\Delta\left({3\over 4}\right)^{1/2}
\left[({\hat k}_{z}{\hat{\bf e}}_{y}-{\hat k}_{y}{\hat{\bf e}}_{z})
+{\rm i}({\hat k}_{x}{\hat{\bf e}}_{z}-{\hat k}_{z}{\hat{\bf e}}_{x})
\right],
\label{F1u1i}
\end{equation}
where ${\hat{\bf k}}\equiv {\bf k}/|{\bf k}|$.  Then, the magnitude of the 
gap is calculated, leading to the expression
\begin{equation}
({\hat{\Delta}}^{\dagger}_{\bf k}{\hat{\Delta}}_{\bf k})_{\pm}=
{3\over 4}\Delta^{2}(1\pm |{\hat k}_{z}|)^{2}.  
\label{eigenvalueF1u1i}
\end{equation}
It is remarked that the amplitude of the smaller gap 
$[({\hat{\Delta}}^{\dagger}_{\bf k}{\hat{\Delta}}_{\bf k})_{-}]^{1/2}$ 
has point nodes in the direction $|{\hat k}_{z}|=1$, and has a quadratic 
dependence as 
$\propto({\hat k}_{x}^{2}+{\hat k}_{y}^{2})$ around the node on the the 
Fermi sphere.  
Therefore, the DOS is proportional to the excitation energy $\omega$.  
With the use of (\ref{eigenvalueF1u1i}), the DOS is calculated numerically 
by means of the formula (\ref{dos}).  The result is shown in 
Fig.\ \ref{fig:1}.  The shape of the DOS is similar to those for the 
polar state.  

The $d$-vector of the state F$_{2u}(1,{\rm i},0)$, the class of group 
theoretical representaion $D_{4}(E)$, is given as~\cite{VG} 
\begin{equation}
{\bf d}({\bf k})=\Delta\left({3\over 4}\right)^{1/2}
\left[
({\hat k}_{z}{\hat{\bf e}}_{y}+{\hat k}_{y}{\hat{\bf e}}_{z})
+{\rm i}({\hat k}_{x}{\hat{\bf e}}_{z}+{\hat k}_{z}{\hat{\bf e}}_{x})
\right]. 
\label{F2u1i}
\end{equation}
The magnitude of the gap is calculated, leading to the same expression as 
(\ref{eigenvalueF1u1i}).  Therefore, the DOS $N_{\rm s}(\omega)$ is the 
same as shown in Fig.\ \ref{fig:1}, the polar-like one.  

\begin{figure}
\begin{center}
\includegraphics[width=0.6\linewidth]{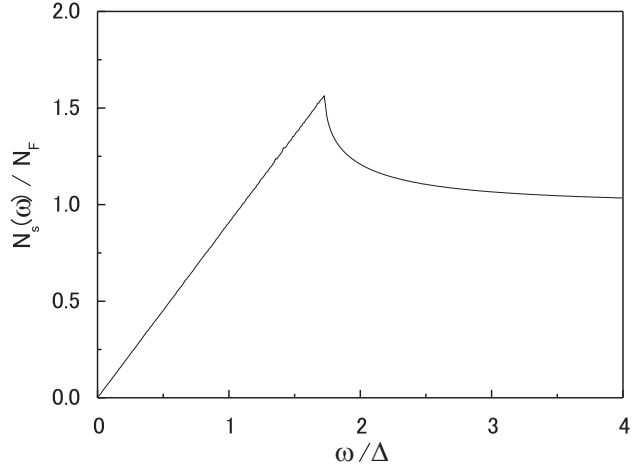}
\caption{
$N_{\rm s}(\omega)/N_{\rm F}$ in the state 
F$_{1{\rm u}}(1,\varepsilon,\varepsilon^{2})$, $N_{\rm F}$ being the DOS 
in the normal state at the Fermi level.  This is the same as that of 
the polar state $\Delta_{\bf k}=\sqrt{3}\Delta{\hat k}_{z}$ within numerical 
errors.  
}
\label{fig:2}
\end{center}
\end{figure}

The $d$-vector of the state F$_{1{\rm u}}(1,\varepsilon,\varepsilon^{2})$, 
the class of group theoretical representaion $D_{3}(E)$, is given as~\cite{VG} 
\begin{equation}
{\bf d}({\bf k})=\Delta\left({1\over 2}\right)^{1/2}
\left[
({\hat k}_{z}{\hat{\bf e}}_{y}-{\hat k}_{y}{\hat{\bf e}}_{z})
+\varepsilon({\hat k}_{x}{\hat{\bf e}}_{z}-{\hat k}_{z}{\hat{\bf e}}_{x})
+\varepsilon^{2}({\hat k}_{y}{\hat{\bf e}}_{x}-{\hat k}_{x}{\hat{\bf e}}_{y})
\right],
\label{F1uee}
\end{equation}
where $\varepsilon\equiv e^{{\rm i}2\pi/3}$.  
Then, the magnitude of the gap is calculated, leading to the expression 
\begin{equation}
({\hat{\Delta}}^{\dagger}_{\bf k}{\hat{\Delta}}_{\bf k})_{\pm}=
{\Delta\over 4}^{2}
\left(\sqrt{3}\pm |{\hat k}_{x}+{\hat k}_{y}+{\hat k}_{z}|\right)^{2}.  
\label{eigenvalueF1uee}
\end{equation}
The smaller gap 
$[({\hat{\Delta}}^{\dagger}_{\bf k}{\hat{\Delta}}_{\bf k})_{-}]^{1/2}$ 
has point nodes in the direction 
${\hat{\bf k}}=(\pm 1/\sqrt{3},\pm 1/\sqrt{3},\pm 1/\sqrt{3})$, and has 
also a quadratic dependence as (\ref{eigenvalueF1u1i}).  
So, the DOS is proportional to the excitation energy $\omega$, and its 
explicit dependence is calculated numerically 
by means of the formula (\ref{dos}).  The result is shown in 
Fig.\ \ref{fig:2}.  The DOS is the same as those for the 
polar state $\Delta_{\bf k}=\sqrt{3}\Delta{\hat k}_{z}$.  

\begin{figure}
\begin{center}
\includegraphics[width=0.6\linewidth]{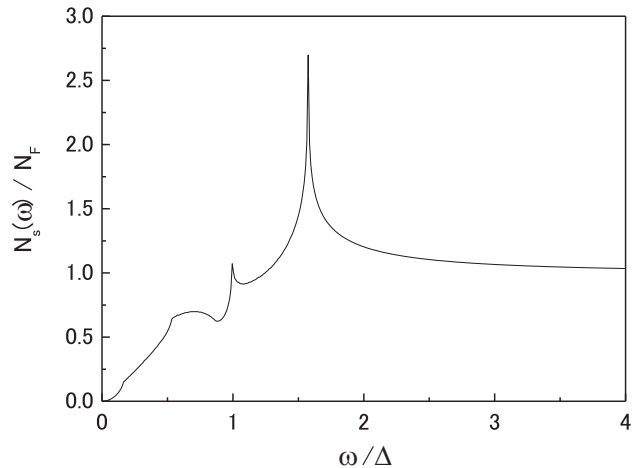}
\caption{
$N_{\rm s}(\omega)/N_{\rm F}$ in the state 
F$_{2{\rm u}}(1,\varepsilon,\varepsilon^{2})$, $N_{\rm F}$ being the DOS 
in the normal state at the Fermi level.  
}
\label{fig:3}
\end{center}
\end{figure}

The $d$-vector of the state F$_{2{\rm u}}(1,\varepsilon,\varepsilon^{2})$, 
the class of group theoretical representaion $D_{3}(E)$, is given as~\cite{VG} 
\begin{equation}
{\bf d}({\bf k})=\Delta\left({1\over 2}\right)^{1/2}
\left[
({\hat k}_{z}{\hat{\bf e}}_{y}+{\hat k}_{y}{\hat{\bf e}}_{z})
+\varepsilon({\hat k}_{x}{\hat{\bf e}}_{z}+{\hat k}_{z}{\hat{\bf e}}_{x})
+\varepsilon^{2}({\hat k}_{y}{\hat{\bf e}}_{x}+{\hat k}_{x}{\hat{\bf e}}_{y})
\right].
\label{F2uee}
\end{equation}
The magnitude of the gap is calculated as 
\begin{eqnarray}
& &
({\hat{\Delta}}^{\dagger}_{\bf k}{\hat{\Delta}}_{\bf k})_{\pm}=
{\Delta^{2}\over 2}\biggl\{
2-({\hat k}_{x}{\hat k}_{y}+{\hat k}_{y}{\hat k}_{z}+{\hat k}_{z}{\hat k}_{x})
\nonumber
\\
& &\qquad
\pm\sqrt{3}\left[1-2({\hat k}_{x}{\hat k}_{y}+{\hat k}_{y}{\hat k}_{z}
+{\hat k}_{z}{\hat k}_{x})+4{\hat k}_{x}{\hat k}_{y}{\hat k}_{z}
({\hat k}_{x}+{\hat k}_{y}+{\hat k}_{z})\right]^{1/2}\biggr\}
\label{eigenvalueF2uee}
\end{eqnarray}
The smaller gap 
$[({\hat{\Delta}}^{\dagger}_{\bf k}{\hat{\Delta}}_{\bf k})_{-}]^{1/2}$ 
has point nodes in the direction 
${\hat{\bf k}}=(\pm 1/\sqrt{3},\pm 1/\sqrt{3},\pm 1/\sqrt{3})$.  
The gap vanishes linearly around the nodes so that the DOS is proportional 
to the square of the energy $\omega$.  However, the region of $\omega$ 
where $N_{\rm s}(\omega)\propto\omega^{2}$ holds is very restricted, 
i.e., $\omega<\Delta/10$, as shown in Fig.\ \ref{fig:3}.  
Therefore, the temperature dependence of physical quantities is hard to 
distinguish from those of the polar-like state.  

Other example, in which the point node(s) give the polar-like DOS, 
is the so-called planar state with E$_{2{\rm u}}$ 
symmetry which is unitary state and was proposed as a candidate of 
that of UPt$_{3}$~\cite{Machida,Sauls}.  
Such a state gives a magnitude of the gap as 
\begin{equation}
|\Delta_{\bf k}|\propto
{\hat k}_{z}[({\hat k}_{x}^{2}-{\hat k}_{y}^{2})^{2}+
4{\hat k}_{x}^{2}{\hat k}_{y}^{2}]^{1/2}.
\label{eigenvalueE2u}
\end{equation}
This gap has point nodes at $|{\hat k}_{z}|=1$ and shows the quardatic 
behaviour around the node on the Fermi surface.  So, the quasiparticles 
around the point nodes should also give the polar-like DOS if there exists 
the Fermi surface around the nodes.  

In conclusion, we have pointed out by explicit calculations that 
the superconducting gap with point nodes in the non-unitary states, 
F$_{1{\rm u}}(1,{\rm i},0)$, 
F$_{1{\rm u}}(1,\varepsilon,\varepsilon^{2})$, 
and F$_{2{\rm u}}(1,{\rm i},0)$, classified by Volovik and Gorkov, 
exhibits the polar-like DOS which is proportional to the excitation 
energy itself rather than its square.  This results arises from the 
fact that the $k$-dependence of the gap around the point nodes is 
quadratic rather than linear as expected in general.  

The author acknowledges Y. Aoki for cummunications urging the author 
to calculate the density of states of various states in the Volovik and 
Gorkov classification scheme.  
This work was supported in part by a Grant-in-Aid for COE Research Program 
(No. 10CE2004) by the Ministry of Education, Culture, Sports, Science and 
Technology.

\end{document}